\documentclass[%
 reprint,
 amsmath,amssymb,
 aps,
pre,
showpacs,
nofootinbib
]{revtex4-1}

\usepackage{graphicx}
\usepackage{dcolumn}
\usepackage{bm}

\def\longrightharpoonup{\relbar\joinrel\rightharpoonup}
\def\longleftharpoondown{\leftharpoondown\joinrel\relbar}

\def\longrightleftharpoons{
  \mathop{
    \vcenter{
      \hbox{
      \ooalign{
        \raise1pt\hbox{$\longrightharpoonup\joinrel$}\crcr
	  \lower1pt\hbox{$\longleftharpoondown\joinrel$}
	  }
      }
    }
  }
}

\begin{document}


\title{Energy Transfer in a Molecular Motor in Kramers' Regime }

\author{K.~J.~Challis}
\author{Michael W.~Jack}%
\affiliation{%
 Scion, 49 Sala Street, Rotorua 3046, New Zealand
}%




\date{\today}

\begin{abstract}

We present a theoretical treatment of energy transfer in a molecular motor described in terms of overdamped Brownian motion on a multidimensional tilted periodic potential.  The tilt acts as a thermodynamic force driving the system out of equilibrium and, for non-separable potentials, energy transfer occurs between degrees of freedom.  For deep potential wells, the continuous theory transforms to a discrete master equation that is tractable analytically.  We use this master equation to derive formal expressions for the hopping rates, drift, diffusion, efficiency and rate of energy transfer in terms of the thermodynamic force.  These results span both strong and weak coupling between degrees of freedom, describe the near and far from equilibrium regimes, and are consistent with generalized detailed balance and the Onsager relations.  We thereby derive a number of diverse results for molecular motors within a single theoretical framework.

\end{abstract}


\maketitle

\section{Introduction}

Biological systems use specialized proteins to convert and utilize chemical energy. These \emph{molecular motors} operate far from equilibrium, with minimal inertia, and in the presence of significant thermal fluctuations \cite{Astumian97, Reimann02, Astumian07}.  Insights into their mechanisms are being provided by single-molecule experiments \cite{Nishiyama02, Itoh04, Rondelez05, Sowa05, Arata09, Bustamante11} and the artificial synthesis of molecules that mimic motor proteins \cite{Hernandez04, Astumian07, Xu08, Hanggi09}. Energy transfer in molecular motors has been described by a variety of stochastic theoretical approaches \cite{Reimann02, Magnasco94, Julicher97, Wang08}. A general theory would unify these treatments and provide an opportunity to clarify fundamental aspects of molecular motor operation.

Theoretical descriptions of molecular motors can be broadly categorized into three types: (i) one-dimensional studies of Brownian motion on asymmetric, often time-dependent, periodic potentials \cite{Reimann02}; (ii) discrete master equation treatments \cite{Julicher97, Fisher99, Lattanzi01, Fisher05, Xing06, Wang08, Kim09}; and (iii) descriptions of Brownian motion on a multidimensional free-energy landscape \cite{Magnasco94, Keller00}.  Type (i) theories build on the Feynman ratchet, a model used to demonstrate the impossibility of fluctuations leading to directed motion at equilibrium  \cite{Reimann02}.  The addition of a linear or time-dependent potential drives the system out of equilibrium and enables directed motion.  Type (ii) theories are based on generalizing discrete master equation treatments of chemical reactions \cite{Fisher99, Fisher05, Wang08}.  In these master equations the ratio between forward and backward kinetic rates is constrained by imposing generalized detailed balance \cite{Lattanzi01, Wang03, Xing05}.  Master equation theories have been used to develop detailed phenomenological models of specific molecular motors \cite{Fisher05, Xing06, Kim09}.  Type (iii) theories are based on the idea that chemical reactions can be described as Brownian motion over a potential barrier \cite{Kramers40}.  This means that both chemical and mechanical coordinates can be incorporated within the same theoretical framework: Brownian motion on a  multidimensional time-independent potential \cite{Magnasco94, Keller00}.  In this approach, energy coupling between degrees of freedom occurs for non-separable potentials.  Type (iii) theories are a candidate for a general theory of energy transfer in molecular motors. 

The continuous diffusion equation for Brownian motion on a multidimensional potential is not analytically tractable in general \cite{Risken89, Seifert08}.  This makes it difficult to connect type (iii) theories with experiments, phenomenological models, and established results from non-equilibrium thermodynamics.  However, analytic solutions can be derived in special cases.  For example, in the case of strong coupling, the multidimensional theory reduces to a one-dimensional description along the coupled coordinate \cite{Magnasco94, VandenBroeck12, Golubeva12}.  Analytic solutions can also be found if the degrees of freedom uncouple in a transformed frame \cite{Golubeva12}.  We recently developed an alternative approach that spans both the regime of strong coupling and the more general weakly-coupled case \cite{Challis13}.  In this treatment, the continuous probability density is expanded in a localized Wannier basis to derive a discrete master equation that is analytically tractable. This is the classical analog of the tight-binding model of quantum mechanics and applies for multidimensional non-separable periodic potentials. 

In this paper we formally connect type (iii) theories with a number of well-established results for molecular motors.  We consider the particular case of overdamped Brownian motion on a multidimensional tilted periodic potential.  Using the tight-binding approach, we expand the continuous theory in the Wannier states of the potential to explicitly transform to a master equation that can be interpreted in terms of infrequent hopping between localized discrete states.  For non-separable potentials, this master equation describes hopping transitions that directly couple different degrees of freedom enabling energy transfer.  We extend our previous treatment of this problem \cite{Challis13} by expanding in the Wannier states of the tilted periodic potential rather than the untilted periodic potential.  This generalizes the validity of the master equation from the weak-tilting regime to Kramers' regime.  We use the master equation to derive a range of formal results for molecular motors.  We show that our results are consistent with well-established non-equilibrium thermodynamics results such as generalized detailed balance \cite{Lattanzi01, Wang03, Xing05} and the Onsager relations \cite{Casimir45,Onsager31}.

This paper is organized as follows.  In Sec.\ \ref{sec:Magnasco} we introduce the continuous theory for diffusion on a multidimensional tilted periodic potential and its applicability to molecular motors.  In Sec.\ \ref{sec:tight_binding} we expand in the Wannier states of the potential to derive a discrete master equation.  In Sec.\ \ref{sec:hopping} we consider the master equation hopping rates and connect with Kramers' escape rate and generalized detailed balance.  In Sec.\  \ref{sec:vel} we derive and force-flux relation, and in Sec.\ \ref{sec:powereff} we derive the power-efficiency trade-off.  In Sec.\ \ref{sec:eig} we consider the eigenvalue spectrum of the master equation and the drift and diffusion.  In Sec.\ \ref{sec:entropy} we determine the entropy generation.  In Sec.\ \ref{sec:ccr} we connect our results with coupled chemical reactions.  We conclude in Sec.\ \ref{sec:conc}.

\section{Continuous Theory for Multidimensional Diffusion \label{sec:Magnasco}}

We consider Brownian motion on a multidimensional potential described by the Smoluchowski equation \cite{Risken89} 
\begin{eqnarray}
\frac{\partial P({\bm r},t)}{\partial t} & = & \mathcal{L}P({\bm r},t),
\label{smoluchowski}
\end{eqnarray}
where $P({\bm r},t)$ is the probability density of finding the system at position $\bm r$ at time $t$.  Each degree of freedom is a generalized coordinate capturing the main conformal motions of the molecules and representing displacements in real space or along reaction coordinates \cite{Kramers40}.  In the overdamped limit of negligible inertia, the evolution operator is defined by
\begin{equation}
 \mathcal{L}= \sum_{j} \frac{1}{\gamma_j} \frac{\partial}{\partial r_j} \left[ k_{B}T\frac{\partial}{\partial r_j}+\frac{\partial V({\bm r})}{\partial r_j}\right],
\end{equation}
where $k_{B}$ is the Boltzmann constant, $T$ is the temperature, $\bm \gamma$ is the friction coefficient that may have a different constant value for each degree of freedom, and $j$ is the coordinate index.

The free-energy potential $V({\bm r})$ has both entropic and mechanical contributions \cite{Keller00} and can, in principle, be determined by single-molecule experiments \cite{Hummer01} or molecular dynamics simulations (e.g., \cite{Aksimentiev04}).  We assume a potential in the form of a periodic part with period $\bm a$ and a linear tilt \cite{Keller00}, i.e.,
\begin{equation}
V({\bm r})=V_{\bm 0}({\bm r})-{\bm f}\cdot {\bm r},
\end{equation}
with
\begin{equation}
V_{\bm 0}({\bm r})=V_{\bm 0}({\bm r}+a_j \hat{\bm r}_j)= V_{\bm 0}(\bm r+\bm a).
\end{equation}
The linear potential drives the system out of thermal equilibrium \cite{Reimann02, Challis13}.  It represents a constant macroscopic thermodynamic force due to an external mechanical force or an entropic force such as a concentration gradient across a membrane or an out-of-equilibrium chemical concentration.  Energy transfer occurs when the force in one coordinate induces drift in another.  This is only possible when the potential $V({\bm r})$ contains a non-separable term \cite{Magnasco94}.  Energy transfer in a two-dimensional tilted periodic potential has been demonstrated numerically \cite{Kostur00}.

The above formalism provides a from-first-principles mathematical framework that encompasses all energy transfer in molecular motors, including energy conversion in cytoskeletal motors, rotary motors such as ATP synthase, and ion pumps.  This theory also provides a  physical picture of a molecular-scale system undergoing Brownian motion on a multidimensional time-independent potential that directs the average behavior of the system enabling energy coupling between degrees of freedom for non-separable potentials.  

\section{Transformation to a Discrete Master Equation \label{sec:tight_binding}}

For the case of deep potential wells, the system is strongly localized around the minima of the potential and it is physically intuitive that the continuous theory can be approximated by a discrete equation.  The transformation from a continuous diffusion equation to a discrete master equation represents a significant simplification of the system dynamics and has been attempted by other authors \cite{Keller00, Ferrando93, Jung95, Lattanzi01, Lattanzi02, Wang03, Xing05}.  In our approach, we expand the continuous theory in a localized Wannier basis.  This treatment is analogous to the tight-binding model of a quantum particle in a periodic potential \cite{Kittel04}.  Our previous treatment of this problem expanded the continuous theory in the Wannier states of the untilted periodic potential \cite{Challis13}.  The untilted basis is useful for weak tilting where the force is a small perturbation to the potential.  Here, we expand in the Wannier states of the tilted periodic potential.  Using the tilted basis extends the validity regime of the tight-binding approach beyond weak tilting. 

The evolution operator ${\cal L}$ is periodic so we invoke Bloch's theorem \cite{Kittel04}. The eigenequation for the Smoluchowski equation (\ref{smoluchowski}) is
\begin{equation}
{\cal L}\phi_{\alpha,\bm k}({\bm r})=-\lambda_{\alpha,\bm k}\phi_{\alpha,\bm k}(\bm r),
\label{Lestates}
\end{equation}
where the eigenfunctions $\phi_{\alpha, \bm k}(\bm r)$ have the Bloch form
\begin{equation}
\phi_{\alpha,{\bm k}}({\bm r})=e^{i{\bm k}\cdot{\bm r}}u_{\alpha,{\bm k}}({\bm r}),
\end{equation}
and $u_{\alpha,{\bm k}}({\bm r})$ has the periodicity $\bm a$ of the periodic potential.  The evolution operator is not Hermitian in general so the eigenvalues $\lambda_{\alpha,\bm k}$ have both a real and imaginary part.  The real part is to be interpreted as a decay rate and is due to the Hermitian component of the operator $\cal L$ so Re$\{\lambda_{\alpha\bm k}\}\geq 0$ \cite{Risken89}.  For weak to moderate forcing, the potential minima are well defined and the eigenvalues separate into bands denoted by the band index $\alpha$.  The wavevector $\bm k$ is confined within the first Brillouin zone and, with periodic boundary conditions at infinity, is continuous.  We construct a biothonormal set from the eigenfunctions of $\cal L$ and its adjoint $\cal L^{\dagger}$ \cite{Risken89}.  The adjoint operator is
\begin{equation}
{\cal L}^{\dagger} = \sum_j \frac{1}{\gamma_j} \left[ k_B T \frac{\partial ^2}{\partial r_j^2} - \frac{\partial V(\bm r)}{\partial r_j} \frac{\partial}{\partial r_j}\right],
\end{equation}
and has the eigenequation
\begin{equation}
{\cal L}^{\dagger} \phi_{\alpha, \bm k}^{\dagger}(\bm r) = -\lambda _{\alpha,\bm k}^{\dagger} \phi_{\alpha,\bm k}^{\dagger} (\bm r),
\end{equation}
where the eigenfunctions $\phi_{\alpha,\bm k}^{\dagger}(\bm r)$ also have the Bloch form.  The eigenfunctions satisfy the orthonormality relation 
\begin{equation}
\int d{\bm r} \ \phi_{\alpha,\bm k}^{\dagger *}(\bm r)  \phi_{\alpha',\bm k'}(\bm r)  = \delta_{\alpha,\alpha'} \delta (\bm k -\bm k ').
\end{equation}
Establishing completeness for a non-Hermitian operator is not straight forward.  For the purpose of this work, we assume the completeness relation \cite{Risken89}
\begin{equation}
\sum_{\alpha} \int_{\cal B } d\bm k \ \phi_{\alpha,\bm k}^{\dagger *}(\bm r) \phi_{\alpha ,\bm k}(\bm r ') = \delta(\bm r-\bm r'),
\label{complete}
\end{equation}
where the integral in Eq.\ (\ref{complete}) is denoted by ${\cal B}$ to indicate that it is over a single Brillouin zone.  
The adjoint eigenvalues can be chosen to satisfy
\begin{equation}
\lambda^{\dagger}_{\alpha,\bm k} = \lambda_{\alpha,\bm k}^*.
\end{equation}
The ground state of the adjoint operator is spatially independent with $\lambda_{0, \bm 0}^{\dagger}=0= \lambda_{0, \bm 0}.$

The eigenfunctions $\phi_{\alpha,\bm k}(\bm r)$ and  $\phi_{\alpha,\bm k}^{\dagger}(\bm r)$ are delocalized over the entire spatial extent of the system.  It is convenient to transform to the localized Wannier states 
\begin{equation}
w_{\alpha,{\bm n}}({\bm r})=D\int_{\mathcal{B}} d{\bm k} \  \phi_{\alpha,{\bm k}}({\bm r})e^{{-i {\bm k}\cdot{\bm A}{\bm n}}},
\label{wannier}
\end{equation}
and
\begin{equation}
w_{\alpha,{\bm n}}^{\dagger}({\bm r})=D\int_{\mathcal{B}} d{\bm k} \  \phi_{\alpha,{\bm k}}^{\dagger}({\bm r})e^{{-i {\bm k}\cdot{\bm A}{\bm n}}},
\label{wannier_dagger}
\end{equation}
where $\bm A$ is a diagonal matrix with $A_{jj}=a_j$, $\bm n$ is a vector of integers, and $D=\prod_{j}(a_{j}/2\pi)$.  The Wannier states are a real, discrete, and biothonormal set.  We expand the probability density as
\begin{equation}
P({\bm r},t)=\frac{1}{D}\sum_{\alpha, {\bm n}}p_{\alpha,{\bm n}}(t)w_{\alpha,{\bm n}}({\bm r}),
\label{probden}
\end{equation}
to transform Eq.\ (\ref{smoluchowski}) to the discrete form 
\begin{equation}
\frac{d p_{\alpha,\bm n}(t)}{dt}=\sum_{\alpha',{\bm n}'} \sigma_{\alpha,\alpha',\bm n, \bm n'} p_{\alpha',\bm n'}(t).
\label{master_eqn_alpha}
\end{equation}
The {\em coupling matrix} is
\begin{eqnarray}
\sigma_{\alpha,\alpha',\bm n, \bm n'}& = & \frac{1}{D} \int  d\bm r \  w_{\alpha,\bm n}^{\dagger} (\bm r) {\cal L} w_{\alpha',\bm n'}(\bm r) \\
& = & \kappa_{\alpha,\bm n-\bm n'}\delta_{\alpha\alpha'}
\label{sigma}
\end{eqnarray}
where $\kappa_{\alpha,\bm n}$ are the Fourier components of the eigenvalues, i.e.,
\begin{equation}
\kappa_{\alpha,\bm n} = -D \int_{\cal B} d\bm k \ \lambda_{\alpha,\bm k}e^{i \bm k \cdot \bm A \bm n}.
\end{equation}
Both the coefficients $p_{\bm n}(t)$ and the coupling matrix are real.  The coupling matrix is diagonal in the band index $\alpha$ so each eigenvalue band evolves independently and the system dynamics can be interpreted in terms of intraband hopping between localized Wannier states.  

The band structure of eigenvalues enables a separation of timescales between the rapidly decaying higher bands and the slowly evolving lowest band governing the long-time behavior of the system.  Retaining only the lowest band and dropping the band supscript for the remainder of the paper, we write the resulting master equation 
\begin{equation}
\frac{d p_{\bm n}(t)}{dt}=\sum_{\bm n'}\left[\kappa_{{\bm n}-{\bm n}'}p_{\bm n '}(t)-\kappa_{{\bm n}'-{\bm n}} p_{\bm n}(t)\right], \label{master_eqn}
\end{equation}
where we have used that $\sum_{\bm n}\kappa_{\bm n} =0$.  If the potential wells of the tilted periodic potential are deep compared to the thermal energy $k_B T$, the Wannier states are well localized.  In this case the hopping rates with small $|\bm n|$ dominate and the summation in Eq.\ (\ref{master_eqn}) need only be extended over nearest neighbors\footnote{Nearest-neighbor sums will also be used in the derivation of Eqs.\ (\ref{drift1}), (\ref{drift2}), and (\ref{lambda}).}.  Furthermore, the Wannier states $w_{\bm n}(\bm r)$ are approximately the Gaussian harmonic oscillator states of the potential minima and the adjoint states are approximately $w_{\bm n}^{\dagger}(\bm r) \propto \exp[V(\bm r)/k_B T] w_{\bm n}(\bm r)$.  Taking the Wannier states to be positive, $p_{\bm n}(t)$ is positive and can be interpreted as the probability that the system is localized in the $\bm n$th potential well.

\section{Hopping Rates \label{sec:hopping}}

One of the main benefits of the tight-binding approach is that the discrete master equation is derived explicitly, providing  expressions for the hopping rates $\kappa_{\bm n}$ in terms of the potential, i.e.,
\begin{equation}
\kappa_{\bm n} = \frac{1}{D} \int d\bm r \ w_{\bm n}^{\dagger}(\bm r) {\cal L} w_{\bm 0}(\bm r).
\label{rate}
\end{equation}
The hopping rate $\kappa_{\bm n}$, and in fact the coupling matrix $\sigma_{\alpha,\alpha',\bm n,\bm n'}$, enables direct coupling between different degrees of freedom when the potential $V(\bm r)$ is non-separable.  In contrast, when the potential $V(\bm r)$ is additively separable in all degrees of freedom, the operator ${\cal L}$ is additively separable, the Wannier states are multiplicatively separable and the hopping rates $\kappa_{\bm n}$ describe only transitions occurring independently in each dimension. The degrees of freedom are then uncoupled and energy transfer can not occur.

The hopping rates (\ref{rate}) depend in general on the particular form of the periodic potential $V_{\bm 0}(\bm r)$ and have a complicated functional dependence on the thermodynamic force $\bm f$.  However, in the regime of deep potential wells, there is a connection between nonequilibrium transport in a tilted periodic potential and Kramers' problem of thermal escape from a potential minimum of a deep bistable potential \cite{Kramers40, vanKampen77, vanKampen78, Caroli79, Caroli80, Hanggi90, Gardiner85}.  This enables a simple approximate tilt dependence of the hopping rates to be derived, as follows.  The physical justification for the master equation (\ref{master_eqn}) closely parallels the physical argument in the derivation of Kramers' relation \cite{vanKampen77, Caroli80}: rapid relaxation within potential wells accompanied by slow transitions between wells.  For deep potential wells, the hopping rates $\kappa_{\bm n}$ for nearest-neighbors, i.e., for $|n_j| = 0$ or $1$, dominate and can be approximated by assuming a double-well potential that matches the full potential in the vicinity of the two relevant minima.  We consider the states with $\bm n = \bm 0$ and $\bm n = \bm m$, where $|m_j| = 0,1$.  The master equation (\ref{master_eqn}) can then be approximated by retaining only terms involving $p_{\bm 0}(t)$ and $p_{\bm m}(t)$.  We write 
\begin{eqnarray}
\frac{dp_{\bm 0}(t)}{dt} & = & \kappa_{ -\bm m} p_{\bm m}(t) -\kappa_{\bm m} p_{\bm 0}(t) \\
\frac{dp_{\bm m}(t)}{dt} & = & \kappa_{\bm m} p_{\bm 0}(t) -\kappa_{-\bm m} p_{\bm m}(t).
\end{eqnarray}
Solving this two-state system gives the two eigenvalues $\lambda_0=0$ and
\begin{equation}
\lambda_1 = \kappa_{\bm m}+\kappa_{-\bm m}.
\label{lambda1}
\end{equation} 
Equation (\ref{lambda1}) shows that, for deep wells, the hopping rates can be determined from the first eigenvalue of the double-well approximation to the potential.  If the most probable path for the transition occurs along a straight line between the minima and contains a single dominant saddle point, the eigenvalue $\lambda_1$ can be determined analytically using the WKB method \cite{Caroli80}.  This gives Kramers' escape rate with the tilt dependence
\begin{equation}
\kappa_{\bm n} =e^{ \alpha_{\bm n} \bm f \cdot \bm A \bm n / k_B T} \kappa_{\bm n}^{\bm 0},
\label{rate_K}
\end{equation}
where $\kappa_{\bm n}^{\bm 0}=\kappa_{-\bm n}^{\bm 0}$ is the rate (\ref{rate}) with $\bm f = \bm 0$ \cite{Challis13}, and in Eq.\ (\ref{rate_K}) we have neglected terms in the exponent that are second order in $\bm f$ \cite{Kolomeisky07, Seifert10, VandenBroeck12, Meng13}.  The loading coefficient $\alpha_{\bm n}$ describes the position of the saddle point between consecutive minima and satisfies $0 \leq \alpha_{\bm n} \leq 1$ and $\alpha_{\bm n}+\alpha_{-\bm n}=1$.  For simplicity, we take $\alpha_{\bm n}=1/2$ (unless otherwise stated).  This choice reduces the possibility of interference between transition paths.  Therefore, for the remainder of this paper, we assume the form
\begin{equation}
\kappa_{\bm n} =e^{ \bm f \cdot \bm A \bm n / 2k_B T} \kappa_{\bm n}^{\bm 0}.
\label{rate_use}
\end{equation}

The hopping rates (\ref{rate_use}) can be used to determine the tilt dependence of the ratio between forward and backward hopping rates, i.e.,
\begin{equation}
\frac{\kappa_{{\bm n}}}{\kappa_{-{\bm n}}}=e^{{\bm f}\cdot {\bm A}{\bm n}/ k_B T}. \label{fluctuation_theorems}
\end{equation}
Equation (\ref{fluctuation_theorems}) is consistent with generalized detailed balance for tilted periodic potentials \cite{Lattanzi01, Wang03, Xing05} and is well known in the context of elementary chemical reactions \cite{Lems03}.  In our treatment, condition (\ref{fluctuation_theorems}) is not imposed as a constraint on the theory but is an analytic result derived from the Smoluchowski equation (\ref{smoluchowski}) in the limit of deep potential wells.  

\section{Force-Flux Relation \label{sec:vel}}  
  
Solving the master equation (\ref{master_eqn}) to determine physical properties of the system provides an opportunity to test the theory against established non-equilibrium thermodynamics results.   In particular, the average rate of hopping is given by the spatial drift 
\begin{equation}
{\bm v}=\frac{d\langle {\bm n}\rangle}{dt}=\sum_{\bm n} {\bm n} \frac{d p_{\bm n}(t)}{dt}.
\label{flux0}
\end{equation}
Using the master equation (\ref{master_eqn}), and the functional form of the hopping rates (\ref{rate_use}), the drift can be determined to be 
\begin{equation}
\bm v = \sum_{\bm n}\bm n \kappa_{\bm n} = \sum_{\bm n} \bm n \kappa_{\bm n}^{\bm 0} e^{\bm f \cdot \bm A \bm n / 2k_B T}.
\label{flux}
\end{equation}
Equation (\ref{flux}) shows the functional dependence of the drift on the thermodynamic force, and vanishes for ${\bm f}=\bm 0$.  Interpreting $X_j = f_j a_j/ T$ as the generalized thermodynamic forces and $v_j$ as the conjugate fluxes, Eq.\ (\ref{flux}) represents a generalized force-flux relation.  Near equilibrium, $|X_j|/k_B\ll 1$ and Eq.\ (\ref{flux}) reduces to 
\begin{equation}
\bm v = \sum_{\bm n} \bm n \kappa_{\bm n}^{\bm 0}  \sum_j \frac{X_j n_j}{k_B}.
\label{flux_eq}
\end{equation}
The components of Eq.\ (\ref{flux_eq}) can be written as 
\begin{equation}
v_j = \sum_{j'}L_{jj'} X_{j'},
\end{equation}
where 
\begin{equation}
L_{jj'}= \sum_{\bm n} n_j n_{j'} \kappa_{\bm n}^0 /k_B=L_{j'j}
\end{equation}
satisfies the Onsager relations \cite{Onsager31,Casimir45}.

In the conceptually simpler two-dimensional case, the force-flux relation (\ref{flux}) becomes
\begin{eqnarray}
v_{x}&=&2\kappa^{\bm 0}_{(1,0)} \sinh(X_x/2k_B) \nonumber \\
& & +2\kappa^{\bm 0}_{(1,1)} \sinh(X_x/2k_B+X_y/2k_B)\label{drift1}\\
v_{y}&=&2\kappa^{\bm 0}_{(0,1)}\sinh(X_y/2k_B )+\nonumber \\
& & 2\kappa^{\bm 0}_{(1,1)} \sinh(X_x/2k_B+X_y/2k_B),\label{drift2} 
\end{eqnarray}
where we have assumed only nearest-neighbor transitions and that $|\kappa^{\bm 0}_{(1,-1)}| \ll |\kappa^{\bm 0}_{(1,0)}|, |\kappa^{\bm 0}_{(0,1)}|, |\kappa^{\bm 0}_{(1,1)}|$.    The hopping rates $\kappa^{\bm 0}_{(1,0)}$ and $\kappa^{\bm 0}_{(0,1)}$ represent transitions occurring independently in each dimension. Identifying $X_z = X_x+X_y$ as the thermodynamic force along the coupled coordinate, the hopping rate $\kappa^{\bm 0}_{(1,1)}$ represents transitions occurring along the coupled coordinate and transferring energy between degrees of freedom.  

\section{Power-Efficiency Trade-Off \label{sec:powereff}}

\begin{figure}[t]
	\centering
		\includegraphics{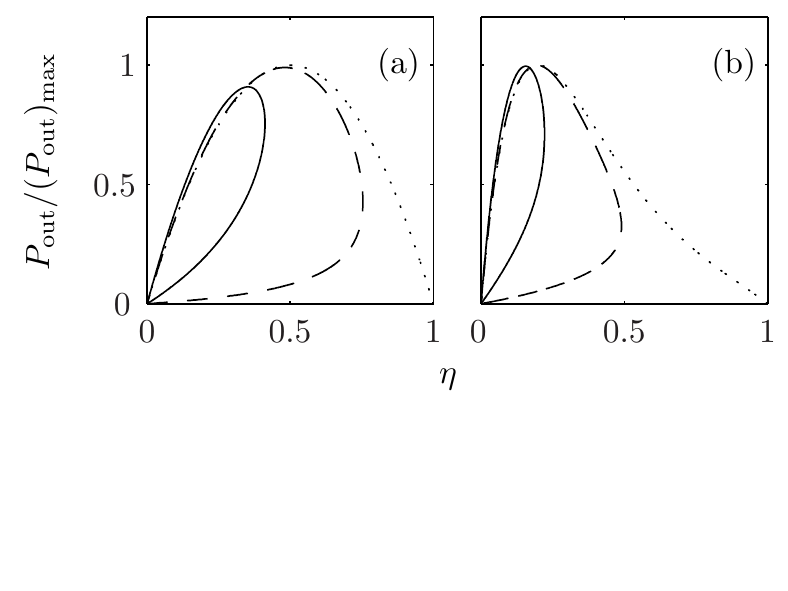}  
	 		\vspace{-2cm}
		\caption{ Normalized output power versus efficiency for $(a)$  $X_{x}/k_B=0.1 $ and $(b)$ $X_x/k_B=10$ with (dotted) $\kappa^{\bm 0}_{(0,1)}=0$, (dashed) $\kappa^{\bm 0}_{(0,1)}=0.01 \kappa^{\bm 0}_{(1,1)}$, and (solid)  $\kappa^{\bm 0}_{(0,1)}=0.1\kappa^{\bm 0}_{(1,1)}$. Other parameters are $\kappa^{\bm 0}_{(1,0)}=\kappa^{\bm 0}_{(0,1)}$. }
	\label{fig:tradeoff}
\end{figure}

Energy transfer processes are characterized by a trade-off between output power and efficiency \cite{Gordon91}.  For a molecular motor, the power-efficiency trade off can be determined from the force-flux relation and may have important biological consequences \cite{Santillan97,Pfeiffer01}.  In the two-dimensional case, we consider that the linear potential is downhill in direction $x$ $(X_x>0)$ and uphill in direction $y$ $(X_y<0)$.  Energy transfer from $x$ to $y$ is thermodynamically viable when the coupling transitions are downhill, i.e., $X_z=X_x+X_y > 0.$  The efficiency of energy transfer can be determined by the ratio of the power output $P_{out}=-v_{y}X_{y}T$ to input $P_{in}=v_{x}X_{x}T$, i.e.,
\begin{equation}
\eta=\frac{P_{out}}{P_{in}}=-\frac{v_{y}X_{y}}{v_{x}X_{x}}.\label{efficiency}
\end{equation} 
Equation (\ref{efficiency}) satisfies $0\leq\eta\leq 1$ and can be written explicitly in terms of $X_{j}$ by inserting Eqs.\ (\ref{drift1}) and (\ref{drift2}).  In the strong coupling regime the independent hopping transitions are negligible, i.e., $|\kappa^{\bm 0}_{(0,1)}|, |\kappa^{\bm 0}_{(1,0)}|\ll |\kappa^{\bm 0}_{(1,1)}|$, and a one-dimensional treatment is possible along the coupled coordinate $X_z$.  In this case, as $X_{x}\rightarrow -X_{y}$, the fluxes vanish, the system approaches thermal equilibrium along the coupled coordinate, and $\eta\rightarrow 1$ \cite{VandenBroeck12}.  The independent transitions due to $\kappa^{\bm 0}_{(1,0)}$ and $\kappa^{\bm 0}_{(0,1)}$ represent dissipative {\em leak processes} that by-pass the coupling mechanism \cite{Lems03,Golubeva12}.  Equation (\ref{efficiency}) can be interpreted as a trade-off between power output $P_{out}$ and efficiency $\eta$.  Figure \ref{fig:tradeoff} shows the power-efficiency trade-off (a) near equilibrium and (b) far from equilibrium.   The dotted lines correspond to the case of strong coupling.  The faster the leak processes the lower the efficiency of the motor.  

\begin{figure}[t]
	\centering
		\includegraphics{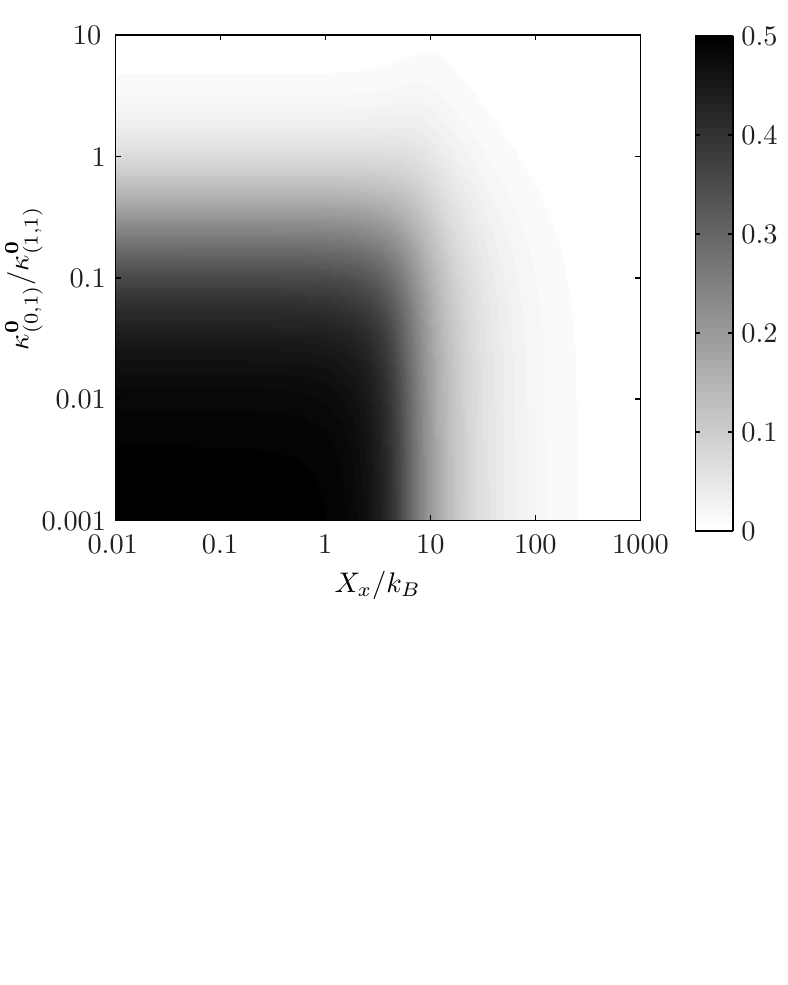}  
			\vspace{-4cm}
		\caption{  Efficiency at maximum power for $\kappa^{\bm 0}_{(1,0)}=\kappa^{\bm 0}_{(0,1)}$.}
	\label{fig:emp}
\end{figure}

The power-efficiency trade off can be used to determine the efficiency at maximum power \cite{Seifert11, Golubeva12, VandenBroeck12}.  The efficiency at maximum power is bounded above by $1/2$ and decreases with increasing rate of the leak processes and with the driving force.  This is shown in Fig.\ (\ref{fig:emp}).  If $\alpha_{\bm n}\neq1/2$, the efficiency at maximum power does not necessarily decrease with increasing driving force and the efficiency at maximum power can exceed $1/2$ \cite{Golubeva12}.  

\section{Eigenvalues, Drift, and Diffusion \label{sec:eig}}

\begin{figure}
	\centering
		\includegraphics{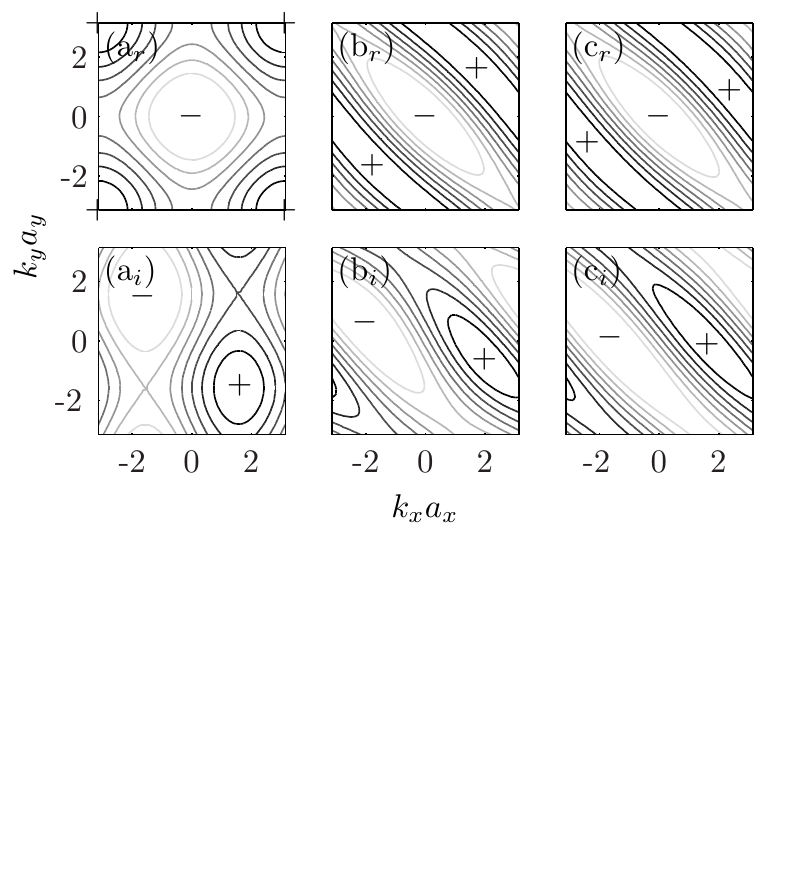}  
		\vspace{-3.8cm}
		\caption{Contour plots of the (upper) real and (lower) imaginary parts of $\lambda_{(k_{x},k_{y})}/\kappa^{\bm 0}_{(0,1)}$ (arbitrary units).  The symbols $+$ and $-$ denote maxima and minima, respectively.  The plots $(a_{r})$ and $(a_{i})$ correspond to  $X_{x}/k_B=0.1 $, $\kappa^{\bm 0}_{(0,1)}=100\kappa^{\bm 0}_{(1,1)}$; $(b_{r})$ and $(b_{i})$ to $X_{x}/k_B=0.1 $, $\kappa^{\bm 0}_{(0,1)}=0.2\kappa^{\bm 0}_{(1,1)}$; and $(c_{r})$ and $(c_{i})$ to $X_{x}/k_B=1$, $\kappa^{\bm 0}_{(0,1)}=0.2\kappa^{\bm 0}_{(1,1)}$. Other parameters are $X_{y}/k_B=-0.05 $ and $\kappa^{\bm 0}_{(1,0)}=\kappa^{\bm 0}_{(0,1)}$. }
	\label{fig:eigenvalues}
\end{figure}

The eigenvalue band structure plays a key role in determining the system properties.  The master equation (\ref{master_eqn}) can be transformed to the diagonal form 
\begin{equation}
\frac{dc_{\bm k}(t)}{dt}=-\lambda_{\bm k}c_{\bm k}(t),
\end{equation}
where the eigenstates are 
\begin{equation}
c_{\bm k}(t) =\sum_{\bm n}p_{\bm n}(t)e^{-i{\bm k}\cdot{\bm A}{\bm n}},
\end{equation}
and the eigenvalues are 
\begin{equation}
\lambda_{\bm k} = - \sum_{\bm n} \kappa_{\bm n} e^{- i \bm k \cdot \bm A \bm n}.
\label{lambdaf}
\end{equation}
Including only nearest-neighbor hopping, Eq.\ (\ref{lambdaf}) can be determined in the two-dimensional case to be
\begin{eqnarray}
\lambda_{(k_x,k_y)}&=&4\kappa^{\bm 0}_{(1,0)}\sin\left(k_{x}a_{x}/2\right)\sin\left(k_{x}a_{x}/2+iX_{x}/2k_B\right)\nonumber\\
&&+4\kappa^{\bm 0}_{(0,1)}\sin\left(k_{y}a_{y}/2\right)\sin\left(k_{y}a_{y}/2+iX_{y}/2k_B\right)\nonumber\\
&&+4\kappa^{\bm 0}_{(1,1)}\sin\left(k_x a_x /2+k_y a_y/2\right)\nonumber \\
& & \times  \sin\left(k_x a_x /2+k_y a_y/2+iX_{z}/2k_B \right). \label{lambda}
\end{eqnarray}
Equation (\ref{lambda}) defines the lowest Bloch band for deep potential wells.  The gradient of the imaginary part at the origin is proportional to the drift, i.e.,
\begin{equation}
{\bm v}\propto \nabla_{\bm k} {\rm Im}(\lambda_{\bm k} ) |_{{\bm k}=\bm 0},
\end{equation}
and the curvature of the real part at the origin is proportional to the time derivative of the covariance matrix \cite{Gardiner85}, i.e., 
\begin{equation}
\frac{d(\langle n_{i}n_{j}\rangle-\langle n_i\rangle\langle n_j\rangle)}{dt}\propto \left. \frac{\partial^{2}{\rm Re}(\lambda_{\bm k})}{\partial k_i\partial k_j} \right|_{{\bm k}=\bm 0}.
\end{equation}  Figure \ref{fig:eigenvalues} shows contour plots of the real and imaginary parts of the eigenvalues throughout the first Brillouin zone for (a) weak coupling near equilibrium, (b) strong coupling near equilibrium, and (c) strong coupling far from equilibrium.  From (a) to (b), the drift goes from $v_{y}<0$ to $v_{y}>0$ despite the fact that $X_{y}<0$.  Only quantitative differences are observed far from equilibrium.

\section{Entropy Generation \label{sec:entropy}}

The entropy of the system in the lowest Bloch band is 
\begin{equation}
S(t) = -k_B \sum_{\bm n} p_{\bm n}(t) \ln p_{\bm n}(t).
\end{equation}
Taking the time derivative yields
\begin{equation}
\frac{dS(t)}{dt}=-k_B \sum_{\bm n} \frac{dp_{\bm n}(t)}{dt} \ln p_{\bm n}(t).
\label{dSdt}
\end{equation}
With the master equation (\ref{master_eqn}), Eq.\ (\ref{dSdt}) can be written as \cite{deGroot84, Seifert05, Tome10, Esposito10}
\begin{equation}
\frac{dS(t)}{dt} = \frac{dS_e(t)}{dt} +\frac{dS_i(t)}{dt},
\end{equation}
where the entropy supplied to the system from the environment is
\begin{equation}
\frac{dS_e(t)}{dt} = -k_B \sum_{\bm n,\bm n'} \kappa_{\bm n'-\bm n} p_{\bm n}(t) \ln \left( \frac{\kappa_{\bm n'-\bm n}}{\kappa_{\bm n-\bm n'}} \right),
\label{dSdt_env}
\end{equation}
and the rate of entropy produced by the system is
\begin{equation}
\frac{dS_i(t)}{dt} = k_B \sum_ {\bm n,\bm n'} \kappa_{\bm n'-\bm n} p_{\bm n} (t) \ln \left( \frac{\kappa_{\bm n'-\bm n}p_{\bm n}(t)}{\kappa_{\bm n-\bm n'} p_{\bm n'}(t)} \right) \geq 0,\label{dSdt_i}
\end{equation}
which is zero for reversible processes and positive for irreversible processes.  Inserting the ratio (\ref{fluctuation_theorems}) of forward to backward hopping rates, and identifying the drift (\ref{flux}), the entropy flow (\ref{dSdt_env}) to the system has the form
\begin{equation}
\frac{dS_e(t)}{dt} = -\frac{\bm f \cdot \bm A \bm v}{T}.
\end{equation}
In the steady state, $p^{ss}_{\bm n}(t)$ is independent of $\bm n$ and $t$, $dS_{ss}(t)/dt = 0$, and the rate of entropy production for the system is
\begin{equation}
\frac{dS^{ss}_i(t)}{dt} = \frac{\bm f \cdot \bm A \bm v}{T}= - \frac{dS_e^{ss}(t)}{dt} \geq 0.
\label{Sss}
\end{equation}
Inserting the generalized thermodynamic forces $X_j=f_j a_j/T$ (see Section \ref{sec:vel}), the rate of entropy production can be written in the familiar form \cite{deGroot84, Julicher97}
\begin{equation}
\frac{dS^{ss}_i(t)}{dt}= \sum_j v_j X_j .
\label{entropy_ss}
\end{equation}

The entropy produced by the system provides a connection to non-equilibrium fluctuation theorems \cite{Astumian07, Crooks99}, as follows.  The form of Eq.\ (\ref{Sss}) suggests that the change of entropy of the system due to a hop by $\bm n$ sites is 
\begin{equation}
\Delta S_{\bm n}=\frac{\bm f \cdot \bm A \bm n}{T}.
\end{equation}
According to the master equation (\ref{master_eqn}), the probability that a hop by $\bm n$ sites occurs within a time $\Delta t$ is given by $\kappa_{{\bm n}} \Delta t$.  The potential is time independent so the time-reverse of that process is a hop by $-\bm n$ sites.  The probability of a backward hop by $\bm n$ sites occuring within a time $\Delta t$ is given by $\kappa_{-{\bm n}} \Delta t$ and the associated change in entropy of the system is $\Delta S_{-\bm n}=-\Delta S_{\bm n}=-\bm f \cdot \bm A \bm n / T.$  Therefore, 
\begin{equation}
\frac{P(\Delta S_{\bm n})}{P(\Delta S_{-\bm n})}=\frac{\kappa_{{\bm n}} }{\kappa_{-{\bm n}} },
\label{probS}
\end{equation}
where $P(\Delta S_{\bm n})=\kappa_{{\bm n}} \Delta t$ is the probability of a hop by $\bm n$ sites occuring in time $\Delta t$ and producing entropy $\Delta S_{\bm n}.$  Using the ratio of forward to backward hopping rates (\ref{fluctuation_theorems}), Eq.\ (\ref{probS}) becomes
\begin{equation}
\frac{P(\Delta S_{\bm n})}{P(\Delta S_{-\bm n})}=e^{\Delta S_{\bm n}/k_B}.
\label{fluc_theo}
\end{equation}
Equation (\ref{fluc_theo}) describes the relative probabilities of discrete hopping events in a form that is consistent with non-equilibrium fluctuation theorems.

\section{Coupled Chemical Reactions \label{sec:ccr}}

To provide a concrete two-dimensional example, consider a coupled chemical reaction system composed of the three elementary reactions \cite{Lems03}
\begin{eqnarray}
A&\mathop{\rightleftharpoons}&B,\quad C \mathop{\rightleftharpoons}D, \quad A+C\mathop{\rightleftharpoons} B+D,\label{reaction}
\end{eqnarray}
numbered 1 to 3 from left to right.  Chemical reactions (at room temperature) can be described via Brownian motion along continuous reaction coordinates but are also often treated as discrete due to the deep potential wells binding the molecules \cite{Hanggi90}.  The thermodynamic forces driving the system are the Gibbs free energies $\Delta G_j$ and thermodynamic consistency requires $\Delta G_{1}+\Delta G_{2}=\Delta G_3$.  The net rate for each chemical reaction is 
\begin{equation}
r_j = R_j^f - R_j^b = R_j^f (1-e^{\Delta G_j/k_B T}),
\label{rxn_rate}
\end{equation}
where $R_j^f$ and $R_j^b$ are the forward and backward reaction rates, respectively, given by the usual mass-action expressions in terms of species activities and reaction rate constants.  In our formalism, the generalized thermodynamic forces are $X_j = -\Delta G_j/ T$ and the generalized fluxes $v_x=r_1+r_3$ and $v_y = r_2+r_3$ are given by the force-flux relations (\ref{drift1}) and (\ref{drift2}).  This is consistent with the reaction rate expressions (\ref{rxn_rate}) and, in addition, predicts the force dependence of the rates $R_j^f$ and $R_j^b$.  As described in Sec.\ \ref{sec:powereff}, if reaction 1 is spontaneous ($\Delta G_{1}<0$), and reaction 2 is non-spontaneous ($\Delta G_2 >0$), reaction 3 enables energy transfer between reactions 1 and 2 and this occurs spontaneously when $\Delta G_3<0$.  
 
In the long-time steady-state, the rate of entropy produced by the system is given by Eq.\ (\ref{entropy_ss}) and can be written as
\begin{equation}
\frac{dS_{ss}(t)}{dt} = \sum_{j=1}^2 v_j X_j = \sum_{j=1}^3 r_j \frac{\Delta G_j}{T}.
\label{Sccr}
\end{equation}
The right-hand side of Eq.\ (\ref{Sccr}) is the sum of the rate of entropy produced for each of the three chemical reactions in Eq.\ (\ref{reaction}) \cite{Lems03}.  Equation (\ref{Sccr}) provides insight into the power and efficiency expressions of Sec.\ \ref{sec:powereff}: the power output is proportional to the entropy produced in the system due to the driven process while the power input is proportional to the entropy produced in the system due to the driving process.  Furthermore, transitions along the coupled coordinate can be interpreted as enabling the thermodynamically spontaneous process to drive the thermodynamically non-spontaneous process. 

\section{Conclusion \label{sec:conc}}

We have described energy transfer in a molecular motor in terms of overdamped Brownian motion on a multidimensional tilted periodic potential. Using a tight-binding approach we derived a discrete master equation valid for long times and deep potential wells.  This master equation is consistent with the Onsager relations and non-equilibrium fluctuation theorems, and predicts a range of other results for molecular motors.  Our approach unifies these results within the single theoretical framework of Brownian motion on a multidimensional free-energy potential.  This framework provides a compelling candidate for a general theory of energy transfer in a molecular motor.

Possible extensions to our work include: (i) detailed comparisons with experiments and phenomenological models; (ii) energy transfer between a tightly bound degree of freedom and a weakly bound one \cite{Reimann02,Keller00}; (iii) multistep systems \cite{Keller00}; (iv) large tilts where long-range hopping transitions occur \cite{Zhang11}; and (v) the inclusion of inertial forces \cite{Risken89}. 

\begin{acknowledgments}
The authors thank S.\ Quesada for preparing power-efficiency trade-off figures.
\end{acknowledgments}


\begin{thebibliography}{99}
\bibitem{Astumian97}
R.\ D.\ Astumian,  
Science {\bf 276}, 917 (1997).
\bibitem{Reimann02}
P.\ Reimann,  
Phys.\ Rep.\ {\bf 361}, 57 (2002).
\bibitem{Astumian07}
R.\ D.\ Astumian,  
Phys.\ Chem.\ Chem.\ Phys.\ {\bf 9}, 5067 (2007).
\bibitem{Bustamante11}
C.\ Bustamante, W.\ Cheng, and Y.\ X.\ Mejia, 
Cell {\bf 144}, 480 (2011).
\bibitem{Nishiyama02}
M.\ Nishiyama, H.\ Higuchi, and T.\ Yanagida, Nature Cell Biology {\bf 4}, 790 (2002).
\bibitem{Itoh04}
H.\ Itoh, A.\ Takahashi, K.\ Adachi, H.\ Noji, R.\ Yasuda, M.\ Yoshida, and K.\ Kinosita, Jr,    
Nature {\bf 427}, 465 (2004).
\bibitem{Rondelez05}
Y.\ Rondelez, G.\ Tresset, T.\ Nakashima, Y.\ Kato-Yamada, H.\ Fujita, S.\ Takeuchi, and H.\ Noji,  
Nature {\bf 433}, 773 (2005).
\bibitem{Sowa05}
Y.\ Sowa, A.\ D.\ Rowe, M.\ C.\ Leake, T.\ Yakushi, M.\ Homma, A.\ Ishijima, and R.\ M.\ Berry,  
Nature {\bf 437}, 916 (2005).
\bibitem{Arata09}
H.\ Arata, A.\ Dupont, J.\ Min\'{e}-Hattab, L.\ Disseau, A.\ Renodon-Corni\`{e}re, M.\ Takahashi, J.-L.\ Viovy, and G.\ Cappello, 
Proc.\ Natl.\ Acad.\ Sci.\ USA {\bf 106}, 19239 (2009).
\bibitem{Hanggi09}
P.\ H\"{a}nggi and F.\ Marchesoni, 
Rev.\ Mod.\ Phys.\ {\bf 81}, 387 (2009).
\bibitem{Xu08}
J.\ Xu and D.\ A.\ Lavan,  
Nat.\ Nanotechnol.\ {\bf 3}, 666 (2008).
\bibitem{Hernandez04}
J.\ V.\ Hern\'{a}ndez, E.\ R.\ Kay, and D.\ A.\ Leigh, 
Science {\bf 306}, 1532 (2004).
\bibitem{Magnasco94}
 M.\ O.\ Magnasco, 
Phys.\ Rev.\ Lett.\ {\bf 72}, 2656 (1994).
\bibitem{Julicher97} 
F.\ J\"ulicher, A.\ Ajdari, and J.\ Prost, Rev.\ Mod.\ Phys.\ {\bf 69}, 1269 (1997).
\bibitem{Wang08} H.\ Wang, J.\ Comput.\ Theor.\ Nanosci.\ {\bf 5}, 2311 (2008).
\bibitem{Fisher99} 
M.\ E.\ Fisher and A.\ B.\ Kolomeisky, Proc.\ Natl.\ Acad.\ Sci.\ USA {\bf 96}, 6597 (1999).
\bibitem{Lattanzi01}
G.\ Lattanzi and A.\ Maritan, Phys.\ Rev.\ E {\bf 64}, 061905 (2001).
\bibitem{Xing06}
J.\ Xing, F.\ Bai, R.\ Berry, and G.\ Oster, 
Proc.\ Natl.\ Acad.\ Sci.\ USA {\bf 103}, 1260 (2006).
\bibitem{Fisher05} M. E. Fisher and Y. C. Kim, Proc.\ Natl.\ Acad.\ Sci.\ USA {\bf 102}, 16209 (2005).
\bibitem{Kim09} Y.\ C.\ Kim, M.\ Wikstr\"{o}m, and G.\ Hummer, Proc.\ Natl.\ Acad.\ Sci.\ USA {\bf 106}, 13707 (2009).
\bibitem{Keller00}
 D.\ Keller and C.\ Bustamante, 
Biophys.\ J.\ {\bf 78}, 541 (2000).
\bibitem{Wang03}
H.\ Wang, C.\ S.\ Peskin, and T.\ C. Elston, J.\ Theor.\ Biol.\ {\bf 221}, 491 (2003).
\bibitem{Xing05}
J.\ Xing, H.\ Wang, and G.\ Oster, Biophys.\ J.\ {\bf 89}, 1551 (2005).
\bibitem{Kramers40}
H.\ A.\ Kramers,  
Physica {\bf 7}, 284 (1940).
\bibitem{Risken89} 
H.\ Risken, \emph{The Fokker-Planck Equation: Methods of Solution and Applications - 2nd ed.} (Springer-Verlag, Berlin, 1989).
\bibitem{Seifert08} U.\ Seifert, Eur.\ Phys.\ J.\ B {\bf 64}, 423 (2008).
\bibitem{VandenBroeck12} C.\ Van den Broeck, N.\ Kumar, and K.\ Lindenberg, Phys.\ Rev.\ Lett.\ {\bf 108}, 210602 (2012).
\bibitem{Golubeva12}
N.\ Golubeva, A.\ Imparato, and L.\ Peliti, EPL {\bf 97}, 60005 (2012).
\bibitem{Challis13}
K.\ J.\ Challis and M.\ W.\ Jack, Phys.\ Rev.\ E {\bf 87}, 052102 (2013).
\bibitem{Onsager31}
L.\ Onsager,  
Phys.\ Rev.\ {\bf 38}, 2265 (1931).
\bibitem{Casimir45}
H.\ B.\ G.\ Casimir,  
Rev.\ Mod.\ Phys.\ {\bf 17}, 343 (1945).
\bibitem{Hummer01}
G.\ Hummer and A.\ Szabo, Proc.\ Natl.\ Acad.\ Sci.\ USA {\bf 98}, 3658 (2001).
\bibitem{Aksimentiev04}
A.\ Aksimentiev, I.\ A.\ Balabin, R.\ H.\ Fillingame, and K.\ Schulten, Biophys.\ J.\ {\bf 86} 1332 (2004).
\bibitem{Kostur00} M.\ Kostur and L.\ Schimansky-Geier, Phys.\ Lett.\ A {\bf 265}, 337 (2000).
\bibitem{Ferrando93}
R.\ Ferrando, R.\ Spadacini, and G.\ E.\ Tommei, Phys.\ Rev.\ E {\bf 48}, 2437 (1993).
\bibitem{Jung95}
P.\ Jung and B.\ J.\ Berne, in \emph{New Trends in Kramers' Reaction Rate Theory}, edited by P.\ Talkner and P.\ H\"{a}nggi,  (Kluwer Academic, Netherlands, 1995), p.\ 67.
\bibitem{Lattanzi02}
G.\ Lattanzi and A.\ Maritan, J.\ Chem.\ Phys.\ {\bf 117}, 10339 (2002).
\bibitem{Kittel04}
C.\ Kittel, {\em Introduction to Solid State Physics }
(Wiley, New York, 2004).
\bibitem{vanKampen77}
 N.\ G.\ van Kampen, 
J.\ Stat.\ Phys.\ {\bf 17}, 71 (1977).
\bibitem{vanKampen78} N.\ G.\ van Kampen, Supplement of the Progress of Theoretical Physics {\bf 64}, 389 (1978).
\bibitem{Caroli79} B.\ Caroli, C.\ Caroli, and B.\ Roulet, J.\ Stat.\ Phys.\ {\bf 21}, 415 (1979).
\bibitem{Caroli80}
B.\ Caroli, C.\ Caroli, B.\ Roulet, and J.\ F.\ Gouyet, J.\ Stat.\ Phys.\ {\bf 22}, 515 (1980).
\bibitem{Hanggi90}
 P.\ H\"{a}nggi, P.\ Talkner,  and  M.\ Borkovec, Rev.\ Mod.\ Phys.\ {\bf 62} 251 (1990).
\bibitem{Gardiner85}
C.\ W.\ Gardiner,  {\em Handbook of Stochastic Methods for Physics, Chemistry and the Natural Sciences - 2nd ed.} (Springer-Verlag, New York, 1985).
\bibitem{Lems03}
S.\ Lems, H.\ J.\  van der Kooi, and J.\ de Swaan Arons, 
Chem.\ Eng.\ Sci.\ {\bf 58}, 2001 (2003).
\bibitem{Kolomeisky07}
A.\ B.\ Kolomeisky and M.\ E.\ Fisher, Annu.\ Rev.\ Phys.\ Chem.\ {\bf 58}, 675 (2007).
\bibitem{Seifert10}
U.\ Seifert, Phys.\ Rev.\ Lett.\ {\bf 104}, 138101 (2010).
\bibitem{Meng13}
X.\ Meng, M.\ Yu, and Y.\ Zhang, J.\ Phys.\ Cond.\ Matt.\ {\bf 25}, 374102 (2013).
\bibitem{Gordon91}
J.\ M.\ Gordon, Am.\ J.\ Phys.\ {\bf 59}, 551 (1991).
\bibitem{Santillan97}
M.\ Santill\'{a}n and F.\ Angulo-Brown, 
J.\ Theor.\ Biol.\ {\bf 189}, 391 (1997).
\bibitem{Pfeiffer01}
T.\ Pfeiffer, S.\ Schuster, and S.\ Bonhoeffer, 
Science {\bf 292}, 504 (2001).
\bibitem{Seifert11} U.\ Seifert, Phys.\ Rev.\ Lett.\ {\bf 106}, 020601 (2011).
\bibitem{deGroot84}
S.\ R.\ de Groot and P.\ Mazur, {\em Non-equilibrium thermodynamics} (Dover, New York, 1984).
\bibitem{Seifert05}
U.\ Seifert, Phys.\ Rev.\ Lett.\ {\bf 95}, 040602 (2005).
\bibitem{Tome10}
T.\ Tom\'{e} and M.\ J. de Oliveira, Phys.\ Rev.\ E {\bf 82}, 021120 (2010).
\bibitem{Esposito10}
M.\ Esposito and C.\ Van den Broeck, Phys.\ Rev.\ E {\bf 82}, 011143 (2010).
\bibitem{Crooks99}
 G.\ E.\ Crooks, 
Phys.\ Rev.\ E {\bf 60}, 2721 (1999).
\bibitem{Zhang11}
Y.\ Zhang, Phys.\ Rev.\ E {\bf 84}, 031104 (2011).




\end{thebibliography}

\end{document}